\newcommand{\ket}[1]{\left|#1\right>}

\newcommand{\inner}[2]{\left(#1|#2\right)}

\newcommand{\nn}{\nonumber\\} 
 
\newcommand{\f}[1]{\mbox{\boldmath$#1$}}

\newcommand{\vau}{\mbox{\boldmath$v$}}
\newcommand{\na}{\mbox{\boldmath$\nabla$}}
\newcommand{\bea}{\begin{eqnarray}}
\newcommand{\ea}{\end{eqnarray}}
\newcommand{\eea}{\end{eqnarray}}
\newcommand{\ord}{{\cal O}}

\def\beq{\begin{equation}}
\def\eeq{\end{equation}}
\def\bea{\begin{eqnarray}} 
\def\eea {\end{eqnarray}}

\documentstyle[aps,prd,twocolumn,amstex,epsfig,psfig]{revtex}

\begin{document}

\wideabs{
\title{On Slow Light as a Black Hole Analogue}
\author{W.~G.~Unruh and R.~Sch\"utzhold}
\address{
Canadian Institute for Advanced Research Cosmology Program,\\ 
Department of Physics and Astronomy, University of British Columbia,\\
Vancouver, British Columbia, Canada V6T 1Z1,\\
email: {\tt unruh@@physics.ubc.ca, schuetz@@physics.ubc.ca}}
\maketitle
\begin{abstract}
Although slow light (electromagnetically induced transparency) would seem 
an ideal medium in which to institute a ``dumb hole'' (black hole analog), 
it suffers from a number of problems. 
We show that the high phase velocity in the slow light regime ensures that 
the system cannot be used as an analog displaying Hawking radiation. 
Even though an appropriately designed slow-light set-up may simulate 
classical features of black holes -- such as horizon, mode mixing, 
Bogoliubov coefficients, etc.\ -- it does not reproduce the related 
quantum effects.
\end{abstract}
}

\section{Introduction}\label{Introduction}

The astonishing ability to slow light to speeds of a few meters per second has
been a striking development in quantum optics, see e.g.~\cite{slow}. 
The idea to use matter systems as analogs \cite{unruh} to the (yet unobserved)
Hawking effect \cite{hawking} for black holes has raised the possibility of 
experimentally testing certain assumptions which enter into those 
calculations, see e.g.~\cite{visser}. 
The dependence of those analogs on the detection of sound waves however causes
problems, as the detection technology for light is much more developed than 
for sound, and finding an optical analog to black holes 
\cite{leo+piw,DBHA,leo-nature,novello} could make the experimental detection 
of the analog for Hawking radiation easier, cf.~\cite{pessi}.

Recently Leonhardt \cite{leo+piw,leo-nature} has suggested that slow light 
systems could be used to create such an analog, but that approach has been 
criticized by one of us \cite{pessi}. 
This paper is an amplification of that criticism, looking in detail at the 
use of slow light in such an analog, and trying to understand in what sense 
slow light could be used to create and analog for black holes, and why, 
despite that analog, it will not create the thermal radiation characteristic 
of the Hawking process.

\section{Description of the Set-Up}\label{Description of the Set-up}

In order to generate slow light, one first chooses an atom with a convenient 
set of atomic transitions, cf.~\cite{slow,leo-slow}.
In particular, a system is chosen with two long lived meta-stable or stable 
states, and with one state which is coupled to these two states via dipole 
electromagnetic transitions ($\Lambda$-system). 
Let us call the two lower meta-stable states $\ket{a}$ and $\ket{b}$. 
The third higher energy state is $\ket{c}$.
The two states $\ket{a}$ and $\ket{b}$ are assumed to have energy
$-\omega_a,~-\omega_b$, and $\ket{c}$ has energy zero and decay constant 
$\Gamma>0$.
(I.e., this higher energy state is assumed to have decay channels other 
than electromagnetic radiation to the $\ket{a}$ and $\ket{b}$ states.)

The electromagnetic field, which we will assume has a fixed polarization,
will be represented by the vector potential $A$ where $E=\partial_t A$ 
(temporal gauge).

\subsection{Effective Lagrangian}\label{Effective Lagrangian}

The effective Lagrangian for this system can be written as
($\hbar=c=1$ throughout)
\bea
\label{lag-all}
L=\int dx\;{\cal L}^A + \sum\limits_j\left(L^\psi_j+L^{A\psi}_j\right)
\,,
\ea
with the usual term governing the dynamics of the electromagnetic field
\bea
\label{lag-A}
{\cal L}^A
=\frac{1}{2}\left[E^2-B^2\right]
=\frac{1}{2}\left[(\partial_t A)^2-(\partial_x A)^2\right]
\,,
\ea
and the Lagrangian of the atomic states
\bea
\label{lag-psi}
L^\psi_j
&=&
i\left(
\psi_{aj}^*\,\partial_t\,\psi_{aj}+
\psi_{bj}^*\,\partial_t\,\psi_{bj}+
\psi_{cj}^*\,\partial_t\,\psi_{cj}
\right)
\nn &&+
\omega_a\,\psi_{aj}^*\psi_{aj}+
\omega_b\,\psi_{bj}^*\psi_{bj}+
i\Gamma\,\psi_{cj}^*\psi_{cj}
\,,
\ea
as well as the interaction term in dipole approximation
\bea
\label{lag-int}
L^{A\psi}_j= 
E(x_j)\left(
\epsilon_a\,\psi_{cj}^*\psi_{aj}+
\epsilon_b\,\psi_{cj}^*\psi_{bj}
\right)
+{\rm h.c.}
\,,
\ea
where $x_j$ is the location of the $j$-th atom. 
Here the $\psi_{\dots j}$ are the amplitudes for the $j$-th particle 
being in the corresponding state
$\ket{{\rm atom}\;j}=\psi_{aj}\ket{a}+\psi_{bj}\ket{b}+
\psi_{cj}\ket{c}$ and $\epsilon_a,~\epsilon_b$ are the associated 
dipole transition amplitudes.

In contrast to the usual set-up, i.e., a strong control beam and a weak
(perpendicular) probe beam, let us assume that there is a strong 
background counter-propagating electromagnetic field
\bea
A_0(t,x)
&=& 
\Omega
\left(
\frac{\cos{\theta}}{\epsilon_a\omega_a}\,e^{i\omega_a(t-x)}+
\frac{\sin{\theta}}{\epsilon_b\omega_b}\,e^{i\omega_b(t+x)}
\right)
+{\rm h.c.}
\,,
\nn
\eea
i.e., at the resonant frequencies of the two transitions. 
The mixing angle $\theta$ controls the relative strength of the 
left- and right-moving beam and $\Omega$ denotes the averaged Rabi
frequency\footnote{Note that $\Omega$ is often defined differently, i.e.,
with an additional factor of two.}.  
For a single beam ($\theta=0$ or $\theta=\pi/2$) $\Omega$ reduces to 
the exact Rabi frequency of that beam.
The fact that the phase velocity is unity (i.e., the light speed) 
prefigures the fact that the effective dielectric constant of the atoms 
is unity at these transition frequencies when the atoms are in the so 
called "dark state", cf.~\cite{slow,leo-slow}.

In the following we shall assume that we can and are making the rotating wave
approximation. 
One solution, the only (up to an overall phase) non-decaying solution, 
for the atoms is
\bea
\label{dark}
\psi_{aj}^0 &=& +e^{i\omega_a (t-x_j)}\sin{\theta} \,,
\nn
\psi_{bj}^0 &=& -e^{i\omega_b (t+x_j)}\cos{\theta} \,,
\nn
\psi_{cj}^0 &=& 0
\,.
\eea
Since the Rabi oscillations between the states $\ket{a}$ and $\ket{c}$
interfere destructively with those between the states $\ket{b}$ and $\ket{c}$
(leading to a vanishing occupation of $\ket{c}$), this solutions is called
a dark state (no spontaneous emission).

\subsection{Linearization}\label{Linearization}

Let us redefine our electromagnetic field such that
\bea
\label{def-phi}
A(t,x) 
&=&
\left(\Omega\,
\frac{\cos{\theta}}{\epsilon_a\omega_a}+\Phi_a(t,x)\right)e^{-i\omega_a (t-x)}
\nn &&
+\left(\Omega\,
\frac{\sin{\theta}}{\epsilon_b\omega_b}+\Phi_b(t,x)\right)e^{-i\omega_b (t+x)}
+{\rm h.c.}
\,,
\eea
where we are going to assume that both $\Phi_b$ and $\Phi_a$ 
are slowly varying functions of time and space (i.e., beat fluctuations).

Furthermore, let us define
\bea
\psi_{aj} &=& (\Psi_{aj}+\sin{\theta})\,e^{i\omega_a(t-x_j)} \,,
\nn
\psi_{bj} &=& (\Psi_{bj}-\cos{\theta})\,e^{i\omega_b(t+x_j)} \,,
\nn
\psi_{cj} &=& \Psi_{cj}
\,,
\eea
where the new variables $\Psi$ are also assumed to be slowly varying.

Substituting into the Lagrangian, retaining only the second order 
terms\footnote{The zeroth-order contributions decouple and the first-order 
terms vanish after an integration by parts, since the background fields solve 
the equations of motion.} 
in the $\Psi,~\Phi_b,~\Phi_a$, using the rotating wave approximation, 
and neglecting time derivatives of $\Phi_b$ and $\Phi_a$ with respect 
to $\omega_a$ and $\omega_b$ we get the effective (approximated) Lagrangian 
for the beat fluctuations
\bea
\label{lin-phi}
{\cal L}^A\simeq
2i\omega_a\,\Phi_a^*(\partial_t+\partial_x)\Phi_a+
2i\omega_b\,\Phi_b^*(\partial_t-\partial_x)\Phi_b
\,,
\ea
and the atomic states
\bea
\label{lin-psi}
L^\Psi_j
&\simeq&
i\left(
\Psi_{aj}^*\,\partial_t\,\Psi_{aj}+
\Psi_{bj}^*\,\partial_t\,\Psi_{bj}+
\Psi_{cj}^*\,\partial_t\,\Psi_{cj}+
\Gamma\,\Psi_{cj}^*\Psi_{cj}
\right)
\nn &&
-i\Omega\left(
\Psi_{cj}^*\Psi_{aj}\cos{\theta}+
\Psi_{cj}^*\Psi_{bj}\sin{\theta}-
{\rm h.c.}\right)
\,,
\ea
as well as the interaction 
\bea
\label{lin-int}
L^{A\Psi}_j
&\simeq&
-i\omega_a\,\epsilon_a\,\sin{\theta}\,\Phi_a(x_j)\Psi^*_{cj}
+i\omega_b\,\epsilon_b\,\cos{\theta}\,\Phi_b(x_j)\Psi^*_{cj}
\nn &&
+{\rm h.c.}
\ea
%

\section{Equations of Motion}\label{Equations of Motion}

The equations of motion for the particle amplitudes can be derived from the
effective Lagrangian
\bea
\label{eom-psi}
\partial_t\Psi_{aj} 
&=& 
-\Omega\,\cos{\theta}\,\Psi_{cj} \,,
\nn
\partial_t\Psi_{bj} 
&=& 
-\Omega\,\sin{\theta}\,\Psi_{cj} \,,
\nn
\partial_t\Psi_{cj} &=& 
\Omega\left(\cos{\theta}\,\Psi_{aj}+\sin{\theta}\,\Psi_{bj}\right)
-\Gamma\,\Psi_{cj}
\nn &&
+\omega_a\,\epsilon_a\,\sin{\theta}\,\Phi_a(x_j)
-\omega_b\,\epsilon_b\,\cos{\theta}\,\Phi_b(x_j)
\,,
\eea
and the equation of motion for the fields $\Phi_a$ and $\Phi_b$ are
\bea
\label{eom-phi}
2\left(\partial_t+\partial_x\right)\Phi_a
&=&
-\epsilon_a\,\sin{\theta}\sum_j\Psi_{cj}\,\delta(x-x_j) \,,
\nn
2\left(\partial_t-\partial_x\right)\Phi_b
&=&
+\epsilon_b\,\cos{\theta}\sum_j\Psi_{cj}\,\delta(x-x_j)
\,.
\eea
Assuming that the particles are sufficiently closely spaced so that there are
many particles in a space of the order of a wavelength of the field, the sum
over $j$ can be replaced by the density of the particles
\bea
\label{density}
2\left(\partial_t+\partial_x\right)\Phi_a
&=&
-\rho(x)\,\epsilon_a\,\sin{\theta}\,\Psi_{c}(x) \,,
\nn
2\left(\partial_t-\partial_x\right)\Phi_b
&=&
+\rho(x)\,\epsilon_b\,\cos{\theta}\,\Psi_{c}(x)
\,.
\eea
%

\subsection{Effective Dispersion Relation}\label{Effective Dispersion Relation}

Assuming harmonic space-time dependence $e^{-i\omega t+i\kappa x}$
of all of the variables, we can solve the equations of motion for the
atomic amplitudes (\ref{eom-psi})
\bea
\Psi_{cj}(\omega) 
&=& 
\left[
\epsilon_a\,\omega_a\,\sin{\theta}\,\Phi_a(\omega,x_j)-
\epsilon_b\,\omega_b\,\cos{\theta}\,\Phi_b(\omega,x_j)
\right]
\nn && \times
\frac{i\omega}{\omega^2-\Omega^2+i\Gamma\,\omega}
\,,
\eea
and inserting this result into Eq.~(\ref{density}) we finally obtain the 
dispersion relation
\bea
\left(\omega+X(\omega)-\kappa\right)
\left(\omega+Y(\omega)+\kappa\right)
=X(\omega)Y(\omega)
\,,
\eea
where
\bea
X(\omega) &=& \frac{\omega}{2}\,
\frac{\rho\,\omega_a\,\epsilon_a^2\,\sin^2{\theta}}
{\Omega^2-\omega^2-i\Gamma\omega}
\,,
\nn
Y(\omega) &=& \frac{\omega}{2}\,
\frac{\rho\,\omega_b\,\epsilon_b^2\,\cos^2{\theta}}
{\Omega^2-\omega^2-i\Gamma\,\omega}
\,.
\eea
%

\subsection{Adiabatic Regime}\label{Adiabatic Regime}

For small $\omega$ and $\kappa$, the dispersion relation derived above
turns out to be linear, i.e., $\omega\propto\kappa$.
Let us specify the required conditions.
As already mentioned above, Eq.~(\ref{density}) is valid for wavelengths
which are much larger than the inter-atomic distance $\Delta x$ 
(typically a few hundreds of nanometers) only
\bea
\label{Delta}
\kappa\ll\frac{1}{\Delta x}
\,.
\ea
In addition, the manipulations of the previous Section
(rotating wave approximation) 
are based on the assumption that the fields $\Phi_b$ and $\Phi_a$
are slowly varying, i.e., $\omega\ll\omega_a,\,\omega_b$.
However, since the Rabi frequency $\Omega$ is supposed to be much 
smaller than the atomic transition energies $\omega_a,\,\omega_b$
and the decay rate is assumed to be small $\Gamma<\Omega$, 
the knee frequency $\Omega$ of the above dispersion relation yields 
the relevant frequency cut-off
\bea
\omega\ll{\rm min}
\left\{\Omega,\,\omega_a,\,\omega_b,\,\frac{\Omega^2}{\Gamma}\right\}
=\Omega
\,.
\ea
In this limit, i.e., in the adiabatic regime, Eq.~(\ref{eom-psi}) 
can be solved via 
\bea
\label{adiabatic}
\Psi_{c}=
\frac{\omega_a\,\epsilon_a\,\sin{\theta}}{\Omega^2}\,\dot\Phi_a-
\frac{\omega_b\,\epsilon_b\,\cos{\theta}}{\Omega^2}\,\dot\Phi_b
\,.
\eea
Rescaling the fields via
\bea
\tilde\Phi_a &=& \omega_a\,\epsilon_a\,\sin{\theta}\,\Phi_a
\,,
\nn
\tilde\Phi_b &=& \omega_b\,\epsilon_b\,\cos{\theta}\,\Phi_b
\,,
\ea
Eqs.~(\ref{eom-phi}) and (\ref{density}) become
\bea
\label{coupling}
\left(\partial_t+\partial_x\right)\tilde\Phi_a &=& - 
\frac{\rho\,\omega_a\,\epsilon_a^2\,\sin^2{\theta}}{2\Omega^2}\,
\left(\partial_t\tilde\Phi_a-\partial_t\tilde\Phi_b\right)
\,,
\nn
\left(\partial_t-\partial_x\right)\tilde\Phi_b &=& +
\frac{\rho\,\omega_b\,\epsilon_b^2\,\cos^2{\theta}}{2\Omega^2}\,
\left(\partial_t\tilde\Phi_a-\partial_t\tilde\Phi_b\right)
\,.
\ea
In order to cast these two first-order differential equations into 
the usual second-order form, let us choose $\theta$ such that\footnote{
Otherwise one would obtain an velocity-like term even for a medium at rest,
cf.~Sec.~\ref{Black Hole Analogue} below. 
However, this term alone cannot generate an effective horizon.} 
\bea
\label{choice}
\frac{\rho\,\omega_a\,\epsilon_a^2\,\sin^2{\theta}}{2\Omega^2}
=
\frac{\rho\,\omega_b\,\epsilon_b^2\,\cos^2{\theta}}{2\Omega^2}
=
\aleph
\,,
\ea
where the dimensionless quantity $\aleph$ describes the slow-down 
of the waves and can be very large $\aleph\gg1$.
In terms of the fields
\bea
\Phi_\pm=\tilde\Phi_a\pm\tilde\Phi_b
\,,
\ea
we can indeed combine the two first-order equalities above
into one second-order equation
\bea
\label{wave}
\left(
\frac{\partial^2}{\partial t^2}-
\frac{\partial}{\partial x}
\frac{1}{1+2\aleph}
\frac{\partial}{\partial x}
\right)\Phi_+=0
\,.
\ea
Obviously, small background fields, i.e., small Rabi frequencies $\Omega$,
may generate a drastic slow-down $\aleph\gg1$.

Note, however, that the above wave equation differs from the equation of 
motion describing a slow-light pulse in the usual set-up -- i.e., a strong 
control beam and a weak (perpendicular) probe beam, cf.~\cite{slow,leo-slow}
\bea
\left([1+\aleph]\partial_t\pm\partial_x\right)\Phi=0
\,.
\ea
Hence the slow-down in Eq.~(\ref{wave}) 
$v_{\rm group}=v_{\rm phase}=1/\sqrt{1+2\aleph}$ 
of the design proposed in the present article is not as extreme as that 
of the usual set-up $v_{\rm group}=1/(1+\aleph) \neq v_{\rm phase}\approx1$,
but still substantial.

\section{Effective Geometry}\label{Effective Geometry}

So far we considered a static medium at rest with a possibly 
position-dependent $\aleph=\aleph(x)$.
Now we allow for a space-time varying variable $\aleph=\aleph(t,x)$,
where the medium is still at rest.
A change of $\aleph$ can be generated by varying $\rho$,
i.e., by adiabatically adding or removing atoms.
The other parameters in Eq.~(\ref{choice}) remain constant -- 
a time-dependent $\Omega$, for example, would generate additional source 
terms and thereby invalidate the background solution.

Furthermore, we shall assume $\omega_a=\omega_b$ as well as 
$\epsilon_a=\epsilon_a$ (which is a reasonable approximation) and hence 
$\theta=\pi/4$ for the sake of simplicity and absorb these quantities by
rescaling the fields $\Phi_\pm$.

\subsection{Effective Action}\label{Effective Action}

Introducing the abbreviation $\f{\Psi}=(\Psi_a,\Psi_b,\Psi_c)^{\rm T}$ the
linearized Lagrangian governing the dynamics of the $\Psi$-fields in 
Eqs.~(\ref{lin-psi}) and (\ref{lin-int}) can be cast into the following form
\bea
{\cal A}^\Psi
&=&
\int d^2x\left(
i\f{\Psi}^\dagger\cdot\dot{\f{\Psi}}+
\f{\Psi}^\dagger\cdot\f{M}\cdot\f{\Psi}
\right.
\nn
&&+
\left.
\left[\f{\Psi}^\dagger\cdot\f{N}\right]\Phi_-+
\left[\f{N}^\dagger\cdot\f{\Psi}\right]\Phi_-^*
\right)
\,,
\ea
with $\f{M}$ denoting a (self-adjoint) $3\times3$ matrix and $\f{N}$ a 
three-component vector as determined by Eqs.~(\ref{lin-psi}) and 
(\ref{lin-int}).
In terms of the differential operator defined via 
$\widehat{\f{D}}=i\partial_t+\f{M}$ and its formal inverse 
$\widehat{\f{D}}^{-1}$ we may complete the square 
\bea
\label{gaussian}
{\cal A}^\Psi
&=&
\int d^2x\left(
\widetilde{\f{\Psi}}^\dagger\cdot\widehat{\f{D}}\cdot\widetilde{\f{\Psi}}-
\Phi_-^*\,\f{N}^\dagger\cdot\widehat{\f{D}}^{-1}\cdot\f{N}\,\Phi_-
\right)
\,,
\ea
with
\bea
\widetilde{\f{\Psi}}=\f{\Psi}+\widehat{\f{D}}^{-1}\cdot\f{N}\,\Phi_-
\,.
\ea
Assuming that the quantum state of the $\Psi$-fields is adequately described 
by the path-integral with the usual (regular) measure ${\mathfrak D}\f{\Psi}$
we are now able to integrate out (i.e., eliminate) those degrees of freedom
explicitly arriving at an effective action for the $\Phi$-fields alone
\bea
\exp\left\{i{\cal A}_{\rm eff}\right\}=
\frac{1}{Z_\Psi}
\int{\mathfrak D}\f{\Psi}\;
\exp\left\{i\left({\cal A}^\Phi+{\cal A}^\Psi\right)\right\}
\,.
\ea
As demonstrated in Eq.~(\ref{gaussian}), the above path-integral is Gaussian 
(${\mathfrak D}\f{\Psi}={\mathfrak D}\widetilde{\f{\Psi}}$) and the associated
Jacobi determinant is independent of $\Phi$.
Hence we obtain
\bea
{\cal A}_{\rm eff}={\cal A}^\Phi-
\int d^2x\;\Phi_-^*\,\f{N}^\dagger\cdot\widehat{\f{D}}^{-1}\cdot\f{N}\,\Phi_-
\,.
\ea
As usual, the inverse differential operator $\widehat{\f{D}}^{-1}$ causes the 
effective action to be non-local (in time) -- but in the adiabatic limit 
$\omega\ll\Omega$, $\aleph\omega\ll\omega_{a,b}$, $\kappa\ll1/\Delta x$, and
$\aleph\kappa\ll\omega_{a,b}$ the low-energy effective action is local
$i\aleph\,\Phi_-^*\dot\Phi_-$.
An easy way to reproduce this result is to remember the original equation of 
motion 
\bea
\widehat{\f{D}}\cdot\f{\Psi}+\f{N}\,\Phi_-=0
\;\leadsto\;
\f{\Psi}=-\widehat{\f{D}}^{-1}\cdot\f{N}\,\Phi_- 
\,,
\ea
and its solution in the adiabatic limit as given by Eq.~(\ref{adiabatic}).
Together with Eq.~(\ref{lin-phi}) we finally arrive at
\bea
\label{A-eff}
{\cal L}_{\rm eff}
&=&
\frac{i}{2}\left(\Phi_+^*\dot\Phi_+
+[1+2\aleph]\Phi_-^*\dot\Phi_-
+\Phi_+^*\Phi_-'
+\Phi_-^*\Phi_+'
\right)
\nn
&&
+{\rm h.c.}
\ea
Strictly speaking, one obtains an effective action for each atom
\bea
{\cal A}_{\rm eff}^j\propto
i\int dt\,\Phi_-^*(t,x^j)\frac{d\Phi_-(t,x^j)}{dt}
+{\rm h.c.}
\,,
\ea
where the total effective action incorporates the sum over all atoms.
With the assumption that the atoms are sufficiently closely spaced, 
cf.~Eq.~(\ref{Delta}), and moving in a direction perpendicular to the beam 
(e.g., in the $y$-direction) only, we recover Eq.~(\ref{A-eff}). 
 
An alternative method for effectively changing the density $\rho$ is to
cause transitions between the states $\ket{a}$ and $\ket{b}$ and further
states $\ket{d}$ and $\ket{e}$, which do not couple to the electromagnetic 
field $\Phi_\pm$ under consideration.
The dynamics of these additional states is governed by the Lagrangian
\bea
{\cal L}_{\rm add}
&=&
i\psi_d^*\dot\psi_d+i\psi_e^*\dot\psi_e+
\omega_d\psi_d^*\psi_d+\omega_e\psi_e^*\psi_e
\nn
&&+
\left(i\dot\zeta\psi_d^*\psi_a+i\dot\zeta\psi_e^*\psi_b+{\rm h.c.}\right)
\,,
\ea
where $i\dot\zeta$ denotes the space-time dependent transition amplitude.
(This particular parameterization will be more convenient later on.) 

If the amplitude (population) of the states $\ket{d}$ and $\ket{e}$ is large
$\psi_{d,e}\gg\psi_{a,b}$ and the transition weak $\zeta\ll1$, we may neglect 
the back-reaction ($\psi_{a,b}\to\psi_{d,e}$) as well as the associated 
(quantum) fluctuations and describe the process by a classical external source
for $\psi_{a,b}$.

Furthermore, assuming $\omega_a=\omega_d$ and $\omega_b=\omega_e$ as well as
\bea
\psi_{d} &=& +e^{i\omega_a (t-x)}\sin{\theta}+\ord(\zeta^2) \,,
\nn
\psi_{e} &=& -e^{i\omega_b (t+x)}\cos{\theta}+\ord(\zeta^2) \,,
\ea
the background solution in Eq.~(\ref{dark}) acquires an overall pre-factor
\bea
\psi_{a}^0 &=& +\zeta(t,x)\,e^{i\omega_a (t-x)}\sin{\theta} \,,
\nn
\psi_{b}^0 &=& -\zeta(t,x)\,e^{i\omega_b (t+x)}\cos{\theta} \,,
\nn
\psi_{c}^0 &=& 0
\,.
\eea
This scale factor $\zeta(t,x)$ enters the subsequent formulas and effectively
changes the density of the contributing atoms $\rho_{\rm eff}=\zeta^2\rho$.
In particular, the wave equation (\ref{coupling}) gets modified via
\bea
\partial_t\Phi_-+\partial_x\Phi_+
+2\aleph\,\zeta\partial_t\left(\zeta\Phi_-\right)=0
\,,
\ea
which is exactly the same equation as derived from the effective action
in Eq.~(\ref{A-eff}) with $\aleph\to\zeta^2\aleph$.

\subsection{Effective Spinor-Representation}\label{Spinor-representation}

According to Eq.~(\ref{A-eff}) the total effective action for the beat 
fluctuations $\Phi_\pm$ of the electromagnetic field can be written as
\bea
\label{eff-A}
{\cal A}
&=&
\frac{i}{2}\int d^2x 
\left[
(1+2\aleph)
\left(\Phi_-^*\partial_t\Phi_- - [\partial_t\Phi_-^*] \Phi_-\right)
\right.
\nn
&&
+
\left(\Phi_+^* \partial_t\Phi_+ - [\partial_t\Phi_+^*] \Phi_+\right)
+\left(\Phi_+^* \partial_x\Phi_- - [\partial_x\Phi_+^*] \Phi_- \right)
\nn
&&
\left.
+\left( \Phi_-^* \partial_x\Phi_+ - [\partial_x\Phi_-^*]\Phi_+ \right)
\right]
\,.
\eea
Introducing the effective two-component spinor ${\f{\psi}}$ 
(not to be confused with the atomic amplitudes $\psi_{a,b,c}$)
\bea
\label{two-component}
{\f{\psi}}=\left(
\begin{array}{c}
\sqrt{1+2\aleph}\,\Phi_- \\
\Phi_+
\end{array}
\right)
\,,
\eea
this action can be rewritten as
\bea
{\cal A}
&=&
\frac{i}{2}\int \frac{d^2x}{\sqrt{1+2\aleph}}
\left[\sqrt{1+2\aleph}\left({\f{\psi}}^\dagger\partial_t{\f{\psi}}
-\left[\partial_t{\f{\psi}}^\dagger\right]{\f{\psi}}\right)
\phantom{\frac{1}{2}}
\right.
\nn
&&
\left.
+
\left({\f{\psi}}^\dagger\sigma_x\partial_x{\f{\psi}} - 
\left[\partial_x{\f{\psi}}^\dagger\right] \sigma_x {\f{\psi}}\right)
-
{\partial_x\aleph\over1+2\aleph}{\f{\psi}}^\dagger i\sigma_y 
{\f{\psi}} \right]
\,,
\nn
\eea
with $\sigma_x,~\sigma_y~,\sigma_z$ being the Pauli (spin) matrices
obeying $\sigma_x^2=\sigma_y^2=\sigma_z^2=\f{1}$.

But this exactly corresponds to the expression for a 1+1 dimensional 
Dirac field ${\f{\psi}}$ 
\bea
\label{Dirac-A}
{\cal A}
&=&
\int d^2x\,\sqrt{-g}\left[\frac{i}{2}\left(
\,\overline{{\f{\psi}}}\gamma^\mu\nabla_\mu{\f{\psi}} - 
\left[\nabla_\mu\overline{\f{\psi}}\,\right]\gamma^\mu{\f{\psi}}
\right)
\right.
\nn 
&&
\left.
\phantom{\frac{1}{2}}
-m\,\overline{\f{\psi}} {\f{\psi}} \right]
\,,
\eea
if we define the Dirac $\gamma$-matrices via
\bea
\label{Dirac-gamma}
\gamma^0 &=& \sqrt{1+2\aleph}\,\sigma_y \,,
\nn
\gamma^1 &=& -i\sigma_z \,,
\eea
and, accordingly, the Dirac adjoint 
($\overline{\f{\psi}}{\f{\psi}},\overline{\f{\psi}}\gamma^\mu{\f{\psi}}
\in{\mathbb R}$)
\bea
\overline{\f{\psi}} ={\f{\psi}}^\dagger \sigma_y
\,,
\eea
as well as introduce the effective mass
\bea
\label{eff-mass}
m = -\frac12\,\frac{1}{1+2\aleph}\,\frac{\partial\aleph}{\partial x}
\,.
\eea
The effective metric is given by $\{\gamma^\mu,\gamma^\nu\}=2g^{\mu\nu}$
\bea
\label{flat}
ds^2=\frac{dt^2}{1+2\aleph} - dx^2
\,,
\eea
and displays the expected slow-down.

For deriving the identity of Eqs.~(\ref{eff-A}) and (\ref{Dirac-A}) we need 
the properties of the spin connection $\Gamma_\mu$ 
(Fock-Ivaneneko coefficient) which enters into the spin derivative
(remember $\partial_\mu(\overline{\f{\psi}}{\f{\psi}})=
(\nabla_\mu\overline{\f{\psi}}){\f{\psi}}+
\overline{\f{\psi}}\nabla_\mu{\f{\psi}}$)
\bea
\nabla_\mu{\f{\psi}}=\partial_\mu{\f{\psi}}
+\Gamma_\mu{\f{\psi}}
\;,\;
\nabla_\mu\overline{\f{\psi}}=\partial_\mu\overline{\f{\psi}}
-\overline{\f{\psi}}\,\Gamma_\mu
\,,
\eea
and is defined by 
($\nabla_\mu(\overline{\f{\psi}}\gamma^\nu{\f{\psi}})=
(\nabla_\mu\overline{\f{\psi}})\gamma^\nu{\f{\psi}}+
\overline{\f{\psi}}\gamma^\nu\nabla_\mu{\f{\psi}}$) 
\bea
\partial_\mu\gamma^\nu +{\Gamma^\nu}_{\rho\mu}\gamma^\rho
=\left[\gamma^\nu,\Gamma_\mu\right] 
\,,
\eea
with ${\Gamma^\nu}_{\rho\mu}$ being the Christoffel symbol.
In our 1+1 dimensional representation, the l.h.s.\ is a linear combination 
of $\sigma_y$ and $\sigma_z$, cf.~Eq.~(\ref{Dirac-gamma}), and, therefore,
the spin connection $\Gamma_\mu$ has to be proportional to $\sigma_x$.
As a result we obtain the relation
\bea
\left\{\Gamma_\mu,\gamma_\nu\right\}=0
\,,
\eea
and thus confirm the identity of Eqs.~(\ref{eff-A}) and (\ref{Dirac-A})
\bea
2i{\cal L}_{m=0}
&=&
\left[\nabla_\mu\overline{\f{\psi}}\,\right]\gamma^\mu{\f{\psi}}-
\overline{\f{\psi}}\gamma^\mu\nabla_\mu{\f{\psi}} 
\nn
&=& 
\left[\partial_\mu\overline{\f{\psi}}\,\right]\gamma^\mu{\f{\psi}}-
\overline{\f{\psi}}\gamma^\mu\partial_\mu{\f{\psi}}
-
\overline{\f{\psi}}\left\{\Gamma_\mu,\gamma^\mu\right\}{\f{\psi}}
\nn
&=&
\left[\partial_\mu\overline{\f{\psi}}\,\right]\gamma^\mu{\f{\psi}}-
\overline{\f{\psi}}\gamma^\mu\partial_\mu{\f{\psi}}
\,.
\eea
Finally, if we were to choose (over some finite region, since $\aleph>0$)
\bea
1+2\aleph(t,x)=f(t)e^{-4mx}
\,,
\ea
the effective mass $m$ in Eq.~(\ref{eff-mass}) would be constant (which, 
however, is not necessary for the introduction of an effective geometry)
and the analogy to the 1+1 dimensional massive Dirac field complete
\bea
\label{Dirac-Eq}
\left(i\gamma^\mu\nabla_\mu-m\right){\f{\psi}}=0 
\,.
\ea
%

\subsection{Effective Energy}\label{Effective Energy}

The energy-momentum tensor of a Dirac field reads
\bea
\label{Dirac-T}
T_{\mu\nu}
&=&
\frac{i}{2}\left(
\,\overline{\f{\psi}}\gamma_{(\mu}\nabla_{\nu)}{\f{\psi}}-
\left[\nabla_{(\mu}\overline{\f{\psi}}\,\right]
\gamma_{\nu)}{\f{\psi}}
\right)
\nn
&=&
\frac{i}{2}\left(
\,\overline{\f{\psi}}\gamma_{(\mu}\partial_{\nu)}{\f{\psi}}-
\left[\partial_{(\mu}\overline{\f{\psi}}\,\right]
\gamma_{\nu)}{\f{\psi}}
\right)
\,,
\ea
where the second equality sign holds in general only in 1+1 dimensions
(in analogy to the simplifications above).
For an arbitrarily space-time dependent $\aleph$, however, there is no energy 
or momentum conservation law associated to this tensor.
But assuming time-translation symmetry as described by the Killing vector
$\xi=\partial/\partial t$ we may construct a conserved energy via
\bea
\label{cons-E}
E=\int d\Sigma_\mu\,T^{\mu\nu}\,\xi_\nu=\int dx\,\sqrt{-g}\;T^0_0
\,,
\ea
which, for the Dirac field in Eq.~(\ref{Dirac-T}) and the metric in 
Eq.~(\ref{flat}), reads
\bea
\label{Dirac-E}
E=\int dx\,\frac{i}{2}\left({\f{\psi}}^\dagger\dot{\f{\psi}}-
\dot{\f{\psi}}^\dagger{\f{\psi}}\right)
\,.
\ea
On the classical level, this quantity is (even in flat space-time) not 
positive definite (as is well-known).
For quantum fields the situation can be different. 
Imposing fermionic (i.e., anti) commutation relations the energy operator
is -- after renormalization of the zero-point energy and definition of the 
vacuum state as the filled Dirac sea -- indeed non-negative
(again in flat space-time).
However, the fields $\f{\psi}=(\sqrt{1+2\aleph}\,\Phi_-,\Phi_+)^{\rm T}$ do not
obey fermionic but bosonic statistics 
(as one would expect, cf.~ Sec.~\ref{Commutation Relations} below) 
and, therefore, the effective energy possesses negative parts.

This fact is not surprising in the context of the electromagnetic field, 
since one has the huge background field with which these perturbations can 
exchange energy.
However, since in the laboratory frame, the background metric is stationary, 
the energy is a conserved quantity, and the potential instability of the 
negative energy will not be triggered.

\subsection{Inner Product}\label{Inner product}

Since the (classical) equation of motion (\ref{Dirac-Eq}) can be described by 
means of an effective metric in Eqs.~(\ref{flat}) and (\ref{PGL}) below, we 
can introduce a conserved inner product for two solutions of the wave equation
${\f{\psi}}_1$ and ${\f{\psi}}_2$. 
As usual, the inner product of the Dirac field can be derived by means of the 
Noether theorem associated to the global $U(1)$-symmetry 
$\f{\psi}\,\to\,e^{i\varphi}\f{\psi}$ and reads
\bea
\inner{{\f{\psi}}_1}{{\f{\psi}}_2}
&=&
\int d\Sigma_\mu\,\overline{\f{\psi}}_1\,\gamma^\mu\,{\f{\psi}}_2
=
\int dx\,\sqrt{-g}\;\overline{\f{\psi}}_1\,\gamma^0\,{\f{\psi}}_2
\nn
&=&
\int dx\,{\f{\psi}}^\dagger_1{\f{\psi}}_2
\,.
\eea
In contrast to the energy in Eq.~(\ref{Dirac-E}), this quantity is non-negative
on the classical level (as well as for quantum fields with bosonic statistics).
If we were to impose fermionic commutation relations, the above pseudo-norm 
would equal the difference of the number of particles and anti-particles and 
hence not be positive definite. 
But for bosonic statistics it is non-negative.

Note that, for a scalar field, the situation is completely different since
in that case, the inner product is not positive definite: 
$\inner{F^*}{F^*}=-\inner{F}{F}$.  

\section{Black Hole Analogue}\label{Black Hole Analogue}

After introducing the notion of the effective geometry we can now design an
analogue of a black or white hole.
To this end we move the medium with a constant velocity $v$ in order to 
be able to correct (tune) the background beam according to the resulting 
Doppler shift.
Again the background solution, i.e., $\Omega$ and $\theta$, should be 
homogeneous if we want to avoid additional source terms for the linearized 
fields.
(In the reference frame of the moving atoms, an inhomogeneous background
becomes time-dependent and thereby also causes a deviation from the dark 
state.)

The only parameter left for influencing the effective geometry is the density
$\rho$. 
In order to arrive at a stationary effective metric, the density profile
in the laboratory frame should be time-independent $\rho=\rho(x)$.
In the rest frame of the fluid, this requirement implies a space-time
dependence of $\rho=\rho(x-vt)$.
At a first glance, such a scenario seems to be inconsistent with a constant 
velocity $v$, but one could arrange a flow profile such as 
$\vau=v\f{e}_x-vy\rho'\f{e}_y/\rho$ which, for a light beam at $y=0$, 
reproduces these properties.
As already mentioned at the end of Sec.~\ref{Effective Action}, an alternative
possibility is to cause transitions $\ket{d},\ket{e}\to\ket{a},\ket{b}$.

Since we are still working with non-relativistic velocities $v\ll1$,
the rest frame of the medium and the laboratory are related by a 
Galilei transformation
\bea
\frac{\partial}{\partial t}\,\to\,
\frac{\partial}{\partial t}+v\,\frac{\partial}{\partial x}
\;,\;
\frac{\partial}{\partial x}\,\to\,\frac{\partial}{\partial x}
\,.
\ea
Having derived a covariant, i.e., coordinate-independent, representation of 
the effective action in Eq.~(\ref{Dirac-A}), this transformation is completely
equivalent to a corresponding change of the effective geometry
\bea
\gamma^0\,\to\,\gamma^0
\;,\;
\gamma^1\,\to\,\gamma^1+v\,\gamma^0
\,.
\ea
The effective metric is then given by the well-known 
Painlev{\'e}-Gullstrand-Lema{\^\i}tre form \cite{PGL} 
\bea
\label{PGL-up}
g^{\mu\nu}_{\rm eff}
=
(1+2\aleph)
\left(
\begin{array}{cc}
1 & v \\
v & v^2-1/(1+2\aleph)
\end{array}
\right)
\,.
\ea
The inverse metric simply reads 
\bea
\label{PGL}
g_{\mu\nu}^{\rm eff}
=
\left(
\begin{array}{cr}
1/(1+2\aleph)-v^2 & v \\
v & -1
\end{array}
\right)
\,.
\ea
Obviously, a horizon ($g_{00}^{\rm eff}=0$) occurs for $v^2=1/(1+2\aleph)$, 
which could be a relatively low velocity and perhaps experimentally accessible.

\subsection{Negative Effective Energy}\label{Negative Effective Energy}

For stationary (in the laboratory frame) parameters $\aleph=\aleph(x)$ and 
$v=\rm const$ one may construct a conserved energy (Noether theorem) of the 
beat fluctuations $\f{\psi}$ via Eq.~(\ref{cons-E}).
Since for a moving medium, the effective metric in Eq.~(\ref{PGL}) has 
off-diagonal elements, the resulting expression is more complicated than
in Eq.~(\ref{Dirac-E}).
For the sake of convenience, we adopt the geometric-optics approximation 
$\omega,\kappa\gg\aleph'/\aleph$ and obtain
\bea
E=\int dx\,
\f{\psi}^\dagger
\,\frac{\omega+
\left(1+\sigma_x\,v\,\sqrt{1+2\aleph}\right)\left(\omega+v\,\kappa\right)}{2}
\,\f{\psi}
\,,
\ea
and, after diagonalization and normalization $\inner{\f{\psi}}{\f{\psi}}=1$,
the solutions for the effective energy assume the following form
\bea
\label{negative}
E_{\rm eff}=\frac{1}{2}\left(\omega+
\left(1\pm v\,\sqrt{1+2\aleph}\right)\left(\omega+v\,\kappa\right)\right)
\,.
\ea
We observe that even the branch $\omega>0$ of the dispersion relation which 
corresponds to a positive energy in flat space-time can become negative beyond 
the horizon $v>1/\sqrt{1+2\aleph}$.
This purely classical phenomenon -- i.e., that the energy measured at 
infinity can become negative beyond the ergo-sphere $g_{00}=0$ --
occurs for real black holes as well and can be considered as the underlying
reason for the mechanism of super-radiance, etc.  

Of course, the total energy of the system as derived by the total action in 
Eq.~(\ref{lag-all}) is always positive. The modes with a negative effective 
energy (pseudo- or quasi-energy, cf.~\cite{stone}) possess a total energy 
which is smaller than that of the background.
In this regard a (classical) mixing of positive and negative (effective) 
energy modes is possible.

\subsection{Bogoliubov Coefficients}\label{Bogoliubov-Dirac}

If the effective metric possesses a horizon, one would expect the usual mixing
of positive and negative energy solutions as governed by the Bogoliubov 
coefficients $\alpha_E$ and $\beta_E$ defined via
\bea
\f{\psi}^{\rm in}_{E}=
\alpha_E\f{\psi}^{\rm out}_{E}+\beta_E\f{\psi}^{\rm out}_{-E}
\,,
\ea
with the positive ($\f{\psi}_{E}$) and negative ($\f{\psi}_{-E}$) energy 
modes, respectively, which are normalized
\bea
\inner{\f{\psi}^{\rm in}_{E}}{\f{\psi}^{\rm in}_{E}}=
\inner{\f{\psi}^{\rm in}_{-E}}{\f{\psi}^{\rm in}_{-E}}
&=&
\nn
\inner{\f{\psi}^{\rm out}_{E}}{\f{\psi}^{\rm out}_{E}}=
\inner{\f{\psi}^{\rm out}_{-E}}{\f{\psi}^{\rm out}_{-E}}
&=&
1
\,,
\ea
and orthogonal
\bea
\inner{\f{\psi}^{\rm in}_{E}}{\f{\psi}^{\rm in}_{-E}}=
\inner{\f{\psi}^{\rm out}_{E}}{\f{\psi}^{\rm out}_{-E}}=0
\,.
\ea
Owing the the positivity of the inner product in Sec.~\ref{Inner product}
the completeness relation has a plus sign instead of a minus as for the scalar
field, i.e.,
\bea
|\alpha_E|^2+|\beta_E|^2=1
\,.
\ea
Consequently we obtain the Fermi-Dirac factor for the scattering
(Bogoliubov) coefficients 
(instead of the Bose-Einstein distribution for scalar fields)
\bea
\label{Fermi-Dirac}
|\beta_E|=e^{\pi E/\kappa_{\rm s}}|\alpha_E|
\;\leadsto\;
|\beta_E|^2=\frac{1}{e^{2\pi E/\kappa_{\rm s}}+1}
\,.
\ea
However, it should be emphasized here that this  mode mixing is {\em a priori}
a purely classical phenomenon and independent of the quantum features 
(commutation relations) -- the fields $\hat{\f{\psi}}$ do not obey Fermi-Dirac
statistics, see the next Section.
Only if the quantum commutation relations assigned a physically reasonable 
particle interpretation to the modes $\f{\psi}_E$ -- as it is the case for a 
truly fermionic Dirac quantum field, for example, but not for the fields 
$\hat\Phi_\pm$ (see below) -- one could infer the (quantum) Hawking radiation.

The surface gravity of the effective horizon at 
$v^2=1/(1+2\aleph)=c_{\rm slow}^2$ depends on the rate of change of the 
velocity of light in the laboratory frame across the horizon
\bea
\kappa_{\rm s}
&=&
\left|
\frac{\partial(|v|-c_{\rm slow})}{\partial x}
\right|_{\rm horizon}
=
\left|
\frac{\partial c_{\rm slow}}{\partial x}
\right|_{\rm horizon}
\nn
&=&
\frac{1}{\sqrt{(1+2\aleph)^3}}\,
\left|
\frac{\partial\aleph}{\partial x}
\right|_{\rm horizon}
\,.
\ea
By comparison with Eq.~(\ref{eff-mass}), we observe that $\kappa_{\rm s}$ 
is of the same order of magnitude as the rest energy induced by the 
effective mass (remember the homogeneous Dirac equation 
$(i\gamma^0\partial_t-m)\f{\psi}=0$)  
\bea
\omega_m
=
\frac{m}{\sqrt{1+2\aleph}}
=
-\frac{1}{2}\,
\frac{1}{\sqrt{(1+2\aleph)^3}}\,
\frac{\partial\aleph}{\partial x}
\,.
\ea
As a result, the relevant mode-mixing effects -- i.e., the Bogoliubov 
$\beta$-coefficients -- are not strongly suppressed by the effective mass.

\section{Commutation Relations}\label{Commutation Relations}

Having derived an effective metric which may exhibit a horizon, 
one is immediately led to the question of whether the system under 
consideration could be used to simulate the Hawking effect.
As it will turn out, the answer is ``no'' -- since the Hawking effect is a 
quantum effect, it is not sufficient to consider the wave equation,
one also has to check the commutation relations which generate the
zero-point fluctuations (the source of the Hawking radiation) according
to the Heisenberg uncertainty principle.
For convenience we shall transform back into the rest frame of the medium 
and assume a constant $\aleph$ for the calculations in this Section.

\subsection{Commutators}\label{Commutators}

Obviously the effective action derived above is intrinsically different from
the one of a charged scalar field, for example.
To make the difference more explicit let us consider the effective 
(adiabatic limit) commutation relations following from Eq.~(\ref{A-eff}).

For any given time $t_0$, the equal-time commutation relations of the
fields $\hat\Phi_\pm$ vanish. Since the equations of motion do not mix
$\hat\Phi_\pm$ with $\hat\Phi_\pm^\dagger$, this remains true for all times 
\bea
\left[\hat\Phi_\pm(t,x),\hat\Phi_\pm(t',x')\right]=
\left[\hat\Phi_\pm^\dagger(t,x),\hat\Phi_\pm^\dagger(t',x')\right]=0
\,.
\ea
According to Eq.~(\ref{A-eff}) the canonical conjugated momenta are
$i\Phi_+^*$ and $i[1+2\aleph]\Phi_-^*$, respectively, and hence we obtain
\bea
\left[\hat\Phi_+(t,x),\hat\Phi_+^\dagger(t,x')\right]=\delta(x-x')
\,,
\ea
and 
\bea
\left[\hat\Phi_-(t,x),\hat\Phi_-^\dagger(t,x')\right]=
\frac{\delta(x-x')}{1+2\aleph}
\,.
\ea
The remaining (equal-time) commutators vanish
\bea
\left[\hat\Phi_+^\dagger(t,x),\hat\Phi_-(t,x')\right]=
\left[\hat\Phi_+(t,x),\hat\Phi_-^\dagger(t,x')\right]=0
\,,
\ea
and the commutation relations for the time-derivatives of the fields can be
inferred from the equations of motion $\dot\Phi_++\Phi_-'=0$ and
$(1+2\aleph)\dot\Phi_-+\Phi_+'=0$.

Remembering the definition of the effective two-component spinor in 
Eq.~(\ref{two-component}) the above relations can be cast into the compact form
\bea
\left[\hat{\f{\psi}}_A(t,x),\hat{\f{\psi}}_B(t',x')\right]=
\left[\hat{\f{\psi}}_A^\dagger(t,x),\hat{\f{\psi}}_B^\dagger(t',x')\right]=0
\,,
\ea
as well as
\bea
\left[\hat{\f{\psi}}_A(t,x),\hat{\f{\psi}}_B^\dagger(t,x')\right]=
\delta_{AB}\delta(x-x')
\,.
\ea
Since the beat fluctuation of the electromagnetic field (coupled to the medium)
do not obey the Pauli exclusion principle, one cannot fill the Dirac sea 
consisting of all negative (effective) energy (in flat space-time) states and 
thereby define a new vacuum state -- as it is possible for fermionic quantum 
fields.

\subsection{Comparison with Other Fields}\label{Comparison}

Let us compare the above commutation relations with those of a 
(1+1 dimensional) Schr\"odinger field $\psi$ 
\bea
\left[\hat\psi(t,x),\hat\psi(t',x')\right]=
\left[\hat\psi^\dagger(t,x),\hat\psi^\dagger(t',x')\right]=0
\,,
\ea
as well as
\bea
\left[\hat\psi(t,x),\hat\psi^\dagger(t,x')\right]=\delta(x-x')
\,,
\ea
on the one hand and and with the commutators of a (1+1 dimensional) 
charged scalar field $\phi$ 
\bea
\left[\hat\phi(t,x),\hat\phi(t',x')\right]=
\left[\hat\phi^\dagger(t,x),\hat\phi^\dagger(t',x')\right]
&=&0
\,,
\nn
\left[\hat\phi(t,x),\hat\phi^\dagger(t,x')\right]=
\left[\hat\phi^\dagger(t,x),\hat\phi(t,x')\right]
&=&0
\,,
\ea
as well as
\bea
\left[\hat\phi(t,x),\partial_t\hat\phi^\dagger(t,x')\right]=i\delta(x-x')
\,,
\ea
on the other hand.
In the latter case (charged scalar field $\phi$), the equation of motion 
can mix positive and negative frequencies and thereby lead to particle 
production -- whereas in the former situation (Schr\"odinger field $\psi$),
the number of particles is conserved.
This difference becomes more evident when one decomposes the fields into
real (self-adjoint) and imaginary (anti-self-adjoint) parts.
For $\psi$, the independent canonical conjugated variables are 
$\Re\psi$ and $\Im\psi$ -- whereas for $\phi$, they are
$\Re\phi$ and $\Re\dot\phi$ (as well as $\Im\phi$ and $\Im\dot\phi$).

Obviously, the commutation relations of the fields $\Phi_\pm$ are clearly 
inconsistent with those of a charged scalar field $\phi$ and show more 
similarity to the (bosonic) Schr\"odinger field.
Therefore, the system under consideration cannot serve as a true analog 
for the quantum effects in the presence of a black hole horizon -- 
such as Hawking radiation -- although it reproduces all classical phenomena.

Since the fields $\hat\Phi_\pm$ describe fluctuations of the electromagnetic 
field, it is also clear that they do not obey the fermionic (anti) commutation 
rules
\bea
\left\{\hat{\f{\psi}}_A(t,x),\hat{\f{\psi}}_B(t,x')\right\}=
\left\{\hat{\f{\psi}}_A^\dagger(t,x),\hat{\f{\psi}}_B^\dagger(t,x')\right\}=0
\,,
\ea
as well as
\bea
\left\{\hat{\f{\psi}}_A(t,x),\hat{\f{\psi}}_B^\dagger(t,x')\right\}=
\delta_{AB}\delta(x-x')
\,.
\ea
An effective Dirac field satisfying bosonic commutation relations might seem
rather strange in view of the spin-statistics theorem.
Indeed, one key ingredient needed for the derivation of this theorem, 
the spectral condition (which is one of the Wightman axioms), is not satisfied
in our case, since the effective energy can become negative owing to the huge
total energy of the background field, see also Sec.~\ref{Effective Energy}.

\subsection{Particle Creation}\label{Particle creation}

In order to answer the question of whether there is any particle creation 
at all in the described slow-light system, one has to clarify the notion 
of (quasi)particles to be created (or not) and to specify the corresponding
(in/out) vacuum state.

For example, an appropriate initial state $\ket{\rm in}$, which is a coherent 
state in terms of the fundamental creation and annihilation operators of 
the electromagnetic field, could be chosen such that it is annihilated by
all fields $\hat\Phi_\pm$, 
\bea
\forall\,x\;:\;
\hat\Phi_+(t_{\rm in},x)\ket{\rm in}=\hat\Phi_-(t_{\rm in},x)\ket{\rm in}=0
\,.
\ea
This is possible because the fields $\hat\Phi_\pm$ are purely
decomposed of positive frequency parts of the electromagnetic field,
i.e., the annihilators, cf.~Eq.~(\ref{def-phi}).
If the effective Hamiltonian of the fields $\hat\Phi_\pm$
(in an asymptotically flat region, i.e., for a homogeneous medium at rest) 
is given by a non-negative bilinear form such as 
\bea
\hat H_{\rm eff}=
\left(\f{\cal D}\f{\hat\psi}\right)^\dagger
\left(\f{\cal D}\f{\hat\psi}\right)
\,,
\ea
with $\f{\cal D}$ denoting a (differential) operator,
the state $\hat\Phi_+\ket{\rm in}=\hat\Phi_-\ket{\rm in}=0$
is indeed the (or at least one) ground state\footnote{
Therefore, it cannot be the equivalent of the Israel-Hartle-Hawking
\cite{IHH} state, in which the Hawking radiation is somewhat hidden by 
the fact that there is no net energy flux.}.

In this case the initial (vacuum) state is annihilated by the
fields $\hat\Phi_\pm$ at all times
\bea
\forall\,t,x\;:\;
\hat\Phi_+(t,x)\ket{\rm in}=\hat\Phi_-(t,x)\ket{\rm in}=0
\,,
\ea
as the time-evolution does not mix $\hat\Phi_\pm$ with 
$\hat\Phi_\pm^\dagger$, and there is no particle creation.

For another initial (vacuum) state (e.g., a squeezed state)
and a different particle concept,
\bea
f\left[\hat\Phi_\pm,\hat\Phi_\pm^\dagger\right]\ket{\rm in'}=0
\,,
\ea
however, some effects of (quasi)particle creation might occur.
These phenomena could be tested by sending in a (multi-mode) 
squeezed state and comparing the number of photons per mode
in the in- and out-states.

Another possible source for (quasi) particle creation is the finite life-time
of the atomic state $\ket{c}$ as represented by the effective decay rate 
$\Gamma$.
Realistically, this decay corresponds to some spontaneous emission process
generated by the quantum fluctuations of the electromagnetic field, 
for example.
Consistent with the fluctuation-dissipation theorem this coupling also 
introduces (quantum) noise, which is not included in our treatment and could
possibly lead to particle creation.
However, this is clearly a pure trans-Planckian effect and cannot be 
interpreted as Hawking radiation.

\section{Dispersion Relation}\label{Dispersion relation}

\begin{figure}[ht]
\centerline{\mbox{\epsfxsize=8cm\epsffile{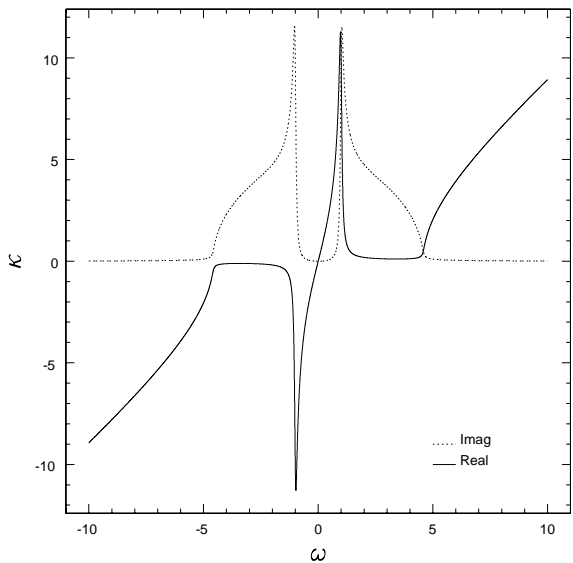}}}
\caption{One branch of the dispersion relation of the $\Phi$-field in 
Eq.~(\ref{disp-slow}).
Frequency $\omega$ and wave-number $\kappa$ are plotted in units of the 
Rabi frequency $\Omega$ for $\aleph=10$ and $\Gamma/\Omega=1/10$.
These values (of order one) are but illustrative and chosen in order to 
resolve the characteristic features in one figure -- realistically the
orders of magnitude are different.
The imaginary part describes the absorption and does not change significantly
in the limit $\Gamma\downarrow0$.
For very large as well as for very small $\omega$ the medium becomes
transparent.
The steep slope within the transparency window $\omega\ll\Omega$ corresponds 
to the reduced propagation velocity -- whereas the effect of the medium for 
large $\omega$ is negligible.
As one can observe, the anomalous frequency solutions $\omega>\Omega$
are separated from the normal ones $\omega<\Omega$ by a large region of 
absorption.}
\label{disp-ko}
\end{figure}

\begin{figure}[ht]
\centerline{\mbox{\epsfxsize=8cm\epsffile{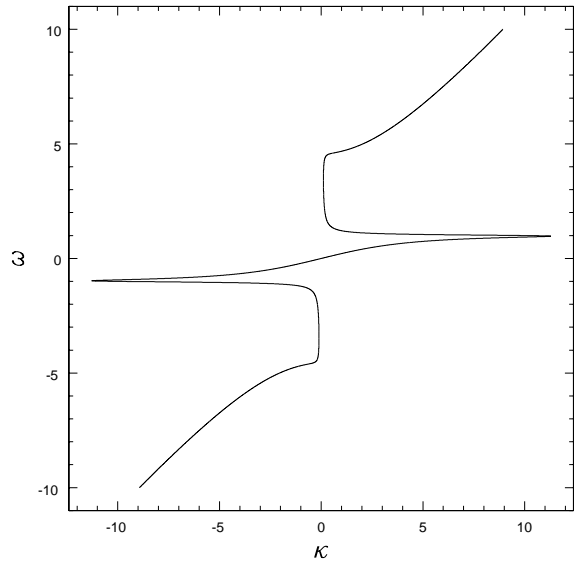}}}
\caption{The real part of the dispersion relation in Fig.~\ref{disp-ko} 
as $\omega$ vs.\ $\kappa$ with the same values.
One can easily recognize that the first deviation from the linear dispersion
relation at $\omega\ll\Omega$ is ``subluminal'' -- although it becomes 
finally ``superluminal'' for $\omega\gg\Omega$.
The solutions with an anomalous (negative or even infinite) group velocity 
lie completely in the absorptive region, cf.~Fig.~\ref{disp-ko}.}
\label{disp-ok}
\end{figure}

Although slow light cannot be used to simulate the Hawking effect it can
reproduce various classical effects associated to 
horizons\footnote{Other systems which are potentially capable of simulating 
those classical effects with present-day technology are discussed in 
Refs.~\cite{gravity} and \cite{volovik}.}, such as mode mixing and the
associated Bogoliubov coefficients, see 
Sec.~\ref{Bogoliubov-Dirac}.
In view of the red- or blue-shift near the horizon deviations from the
linear dispersion relation have to be taken into account, cf.~\cite{jacobson}.
With the choice in Eq.~(\ref{choice}) the dispersion relation in 
Sec.~\ref{Effective Dispersion Relation} simplifies because of 
$X(\omega)=Y(\omega)$, and we obtain for a medium at rest,
cf.~Figs.~\ref{disp-ko} and \ref{disp-ok}
\bea
\label{disp-slow}
\kappa=\pm\omega
\sqrt{1+2\aleph\,\frac{\Omega^2}{\Omega^2-\omega^2-i\Gamma\omega}}
\,.
\ea
%

\begin{figure}[ht]
\centerline{\mbox{\epsfxsize=8cm\epsffile{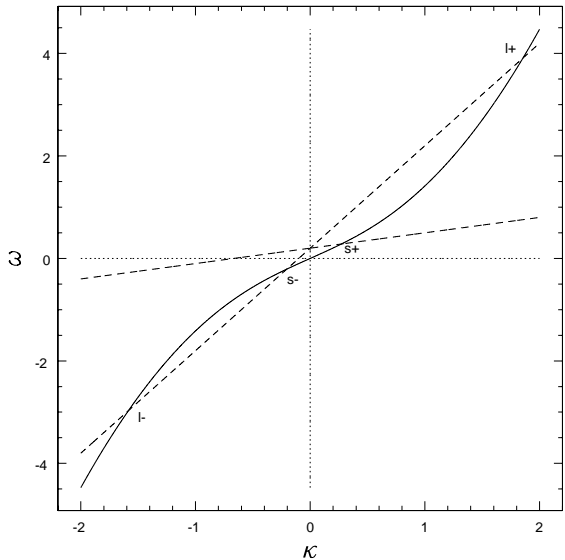}}}
\caption{One branch of the dispersion relation of (zero) sound waves 
in Bose-Einstein condensates at rest, cf.~Eq.~(\ref{disp-sonic}), 
in arbitrary units.
If the condensate is moving the various $\kappa$-solutions for a given
frequency $\omega$ in the laboratory frame can be found by the points of 
intersection with straight lines as determined by Eq.~(\ref{lab-frame}).
For a subsonic velocity $v<c_{\rm sound}$, there is only one solution, 
denoted by s+, which has a small wave-number and a positive pseudo-norm, 
i.e., a positive $\omega_{\rm fluid's\;rest-frame}$
(assuming $\omega_{\rm lab-frame}>0$).
For supersonic velocities, on the other hand, i.e., beyond the horizon,
there are three possible solutions -- one with a small wave-number and a 
negative pseudo-norm (s-) as well as two others with large wave-numbers and 
positive (l+) and negative (l-) pseudo-norm, respectively.
The mixing between these modes at the horizon generates the Hawking
radiation (s+).}
\label{disp-bec}
\end{figure}

We observe two major differences between the dispersion relation above and 
that for the sonic black hole analogs, for example in Bose-Einstein 
condensates (see \cite{BEC} and Sec.~\ref{Bose-Einstein Condensates}) with
\bea
\label{disp-sonic}
\omega^2=c^2_{\rm sound}\kappa^2\left(1+\xi^2\kappa^2\right)
\,,
\ea
where $\xi$ denotes the so-called healing length and provides a 
wave-number cut-off, cf.~Fig.~\ref{disp-bec}.

The sonic black hole analogs generate a deviation from the linear dispersion
relation via the spatial dependence ($\kappa$) and, consequently, for each
value of the wave-number $\kappa$ there exist two possible solutions for the 
frequency ($\pm\omega$ for a medium at rest).
In contrast, for the black hole analogs based on slow light the deviation is 
mainly\footnote{Of course, the finite interatomic distance results in a 
deviation from the linear dispersion relation too, but the cut-off given by
the Rabi frequency is usually reached earlier.} 
caused by the (non-local) temporal dependence.
(This remains true for all dielectric/optical  black hole analogs, 
cf.~\cite{DBHA,novello}.)
As a result, one has two values of $\kappa$ for each value of $\omega$, but 
can have more than two solutions for $\omega$ for some values of $\kappa$. 
Even though these anomalous solutions for $\omega$ are separated from the
normal ones by a relatively large region of absorption, it would be 
interesting to see under which circumstances this peculiar behavior may 
give raise to additional effects (such as mode mixing, etc.).

Another major difference between the dispersion relations (\ref{disp-slow}) 
and (\ref{disp-sonic}) is that the sonic dispersion relation 
(\ref{disp-sonic}) is ``superluminal''/supersonic for large wave-numbers
$v_{\rm group}=d\omega/d\kappa>c_{\rm sound}$ for $\xi\kappa\not\ll1$
whereas the slow-light dispersion relation (\ref{disp-slow})
is ``subluminal'' $v_{\rm group}=d\omega/d\kappa<1/\sqrt{1+2\aleph}$
within the transparency window, say $|\omega|<\Omega/2$, 
but $|\omega|\not\ll\Omega$.
For very large frequencies $\omega\gg\Omega$ one recovers the speed of light 
in vacuum $\omega=\kappa$ -- although this limit is totally outside the region
of applicability of our approximations.

\section{Problems of Slow Light}\label{Problems of Slow Light}

The direct (naive) way to use the most common set-up for slow-light 
experiments -- i.e., a strong control beam and a weak (perpendicular) probe 
beam -- in order to build a black hole analog goes along with a number of 
(somewhat related) difficulties listed below.
Whereas the first three obstacles are can be avoided by the arrangement 
proposed in this article, the fourth one persists -- indicating that 
this system is a classical, but not a quantum analogue of a black hole.

\begin{figure}[ht]
\centerline{\mbox{\epsfxsize=8cm\epsffile{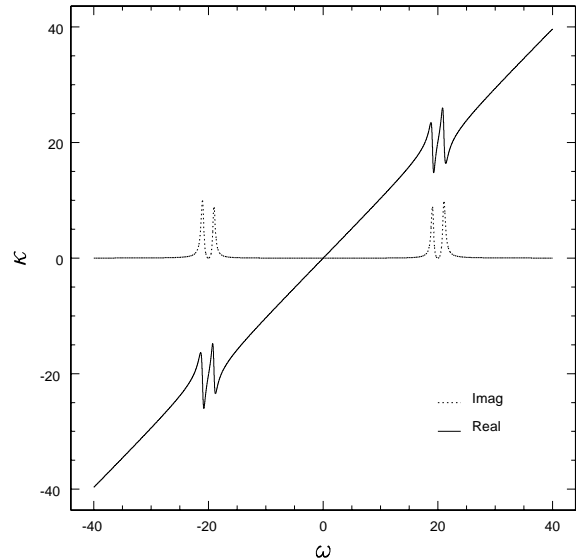}}}
\caption{One branch of the dispersion relation of a slow-light pulse 
(in the usual set-up) 
$\kappa^2=\omega^2[1+\wp(\omega+\omega_0)-\wp(\omega-\omega_0)]$ where 
$\wp(\omega)=2\aleph(\Omega^2/\omega_0)\omega/
(\omega^2-\Omega^2+i\Gamma\omega)$, see e.g., [1,10],
in units of the Rabi frequency $\Omega$ for 
$\omega_0/\Omega=20$, $\Gamma/\Omega=1/2$, and $\aleph=5$.
Again, these unrealistic values have been chosen in order to illustrate 
the chracteristic features.
For more realistic values the peaks would be more pronounced,
the transpacency windows $|\omega\pm\omega_0|\ll\Omega$ narrower,
and the slope inside them steeper, etc., but the main structure remains.
For $|\omega\pm\omega_0|\gg\Omega$ the influence of the medium is negligible.
Within the transpacency windows $|\omega\pm\omega_0|\ll\Omega$, the steep slope
indicates a reduced group velocity and the solutions with an anomalous
group velocity $|\omega\pm\omega_0|=\ord(\Omega)$ lie inside the absorptive 
regions.}
\label{disp-leo-ko}
\end{figure}

\subsection{Frequency Window}\label{Frequency Window}

Light pulses (of the probe beam) are only slowed down drastically 
-- or may propagate at all -- in an extremely narrow frequency window 
in the optical or near-optical regime.
But the frequency of the particles constituting the Hawking radiation
cannot be much larger than the surface gravity
(e.g., the gradient of the fluid's velocity) which makes an experimental
verification in this way very unlikely.

\subsection{Doppler Shift}\label{Doppler Shift}

In a stationary medium, the frequency as measured in the laboratory frame is
conserved -- but the frequency in the atom's rest frame changes as soon as
the velocity of the medium (Doppler shift) or the wave-number (red-shift) 
varies (which necessarily happens near the horizon). 
Hence the beam will leave the narrow frequency window -- which is generated 
by the (moving) atoms -- in general.

\subsection{Group and Phase Velocity}\label{Group and Phase Velocity}

Since the group and the phase velocity of the probe beam are extremely 
different $v_{\rm group} \ll v_{\rm phase} \approx 1$, it is not possible 
to describe its dynamics by an effective local wave equation resembling a 
scalar field in a curved space-time.

\begin{figure}[ht]
\centerline{\mbox{\epsfxsize=8cm\epsffile{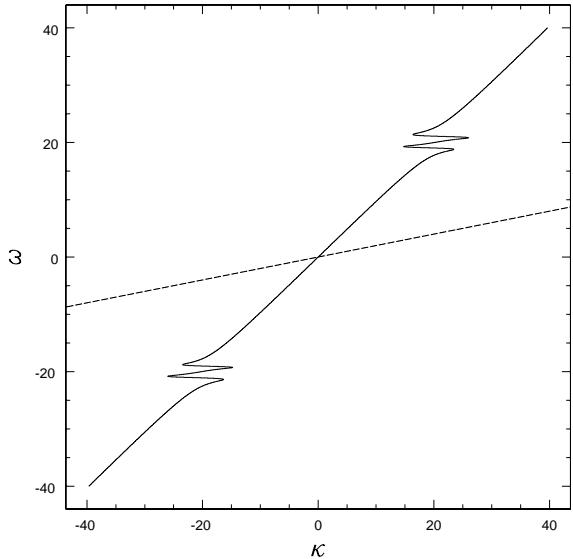}}}
\caption{The real part of the dispersion relation in Fig.~\ref{disp-leo-ko} 
as $\omega$ vs.\ $\kappa$ with the same values.
The additional line demonstrates the slope corresponding to a motion of the 
medium with the reduced group velocity as in Fig.~\ref{disp-bec}. 
Obviously, there can be no mixing of positive and negative pseudo-norms 
via the usual mechanism sketched below Fig.~\ref{disp-bec} in this case. 
Even though the peaks can be much higher for small $\Gamma$ and thereby could
possibly intersect with the straight line, the resulting solutions would lie 
completely in the region of strong absorbtion (cf.~Fig.~\ref{disp-leo-ko}) and
therfore do certainly not model Hawking radiation.}
\label{disp-leo-ok}
\end{figure}

\subsection{Positive/Negative Frequency-Mixing}
\label{Positive/Negative Frequency-Mixing}

In order to obtain particle creation, one has to have a mixing of positive 
and negative frequencies, or, more accurately, positive and negative
pseudo-norm 
(as induced by the inner product, cf.~Sec.~\ref{Bogoliubov-Dirac}) solutions.
In a stationary flowing medium (as used for the black hole analogs),
this can occur by tilting the dispersion relation due to the
Doppler effect caused by the velocity of the medium 
\bea
\label{lab-frame}
\omega_{\rm lab-frame}=\omega_{\rm fluid's\;rest-frame}+v_{\rm medium}\,\kappa 
\,.
\ea
As soon as the velocity on the medium exceeds $|\omega/\kappa|$, 
i.e., the phase velocity, a mixing of positive and negative
frequencies (in the fluid's rest frame) becomes possible, cf.~\cite{pessi}.
However, since the phase velocity of the slow-light pulse is basically
the same as in vacuum, this mechanism does not work in this situation and,
consequently, there is no particle creation.

\section{Comparison with Other Systems}\label{Comparison with other Systems}

One of the main points of the present article is the observation that an 
appropriate wave equation and the resulting effective geometry of a black 
hole analog is {\em not} enough for predicting Hawking radiation.
Although all the classical effects can be reproduced in such a situation,
the adequate simulation of the {\em quantum} effects requires the correct
commutation relations as well.

In view of this observation one might wonder whether this is actually the
case for the currently discussed 
(e.g., sonic/acoustic and dielectric/optical) black hole analogs.
In the following we shall deal with this question for two representative 
examples, for which the commutation relations can be derived easily.

\subsection{Bose-Einstein Condensates}\label{Bose-Einstein Condensates}

The dynamics of Bose-Einstein condensates are to a very good approximation 
described by the Gross-Pitaevskii equation
\bea
i\dot\psi=\left(-\frac{\na^2}{2m}+V(\f{r})+\eta^2|\psi|^2\right)\psi
\,,
\ea
where $\psi$ denotes the mean-field amplitude, $m$ the mass of the bosons,
$V$ an external (trapping) potential, and $\eta$ is the scattering parameter
governing the two-body repulsion of the constituents.
Inserting the eikonal ansatz (Madelung representation),
\bea
\psi=\sqrt{\varrho}\,e^{iS}
\,,
\ea
and introducing the (mean-field) velocity $\f{v}=\na S/m$, one obtains
the equation of continuity $\dot\varrho+\na(\varrho\vau)$ and the
equivalent of the Bernoulli or the Hamilton-Jacobi equation 
\bea
\label{bernoulli}
\dot S+V+\eta^2\varrho+\frac{(\na S)^2}{2m}=
\frac{1}{2m}\,\frac{\na^2\sqrt{\varrho}}{\sqrt{\varrho}}
\,.
\ea
Within the Thomas-Fermi approximation, one neglects the quantum potential, 
i.e., the term on the l.h.s., and hence recovers the usual equations of 
fluid dynamics, see also \cite{BEC}.
The linearization around a given (stationary) background profile 
$\varrho_0$ and $S_0\to\vau_0$ yields the well-known wave equation
\bea
\left(\partial_t+\na\cdot\vau_0\right)
\left(\partial_t+\vau_0\cdot\na\right)\delta S
=
\frac{\eta^2}{m}\,\na\varrho_0\na\delta S
\,.
\ea
The commutation relations of $\delta S$, which we are interested in,
can be derived from the commutator of the fundamental fields
\bea
\left[\hat\psi(t,\f{r}),\hat\psi^\dagger(t,\f{r'})\right]=
\delta^3(\f{r}-\f{r'})
\,.
\ea
Inserting the linearization of $\hat\psi=\sqrt{\hat\varrho}\exp(i\hat S)$
around a classical background via $\hat\varrho=\varrho_0+\delta\hat\varrho$ 
and $\hat S=S_0+\delta\hat S$ we obtain
(note that $\hat\varrho=\hat\varrho^\dagger$ and $\hat S=\hat S^\dagger$)
\bea
\left[\delta\hat\varrho(t,\f{r}),\delta\hat S(t,\f{r'})\right]=
i\delta^3(\f{r}-\f{r'})
\,.
\ea
The relation between $\delta\hat\varrho$ and $\delta\hat S$ follows from
Eq.~(\ref{bernoulli}) in the Thomas-Fermi approximation
\bea
\delta\hat\varrho=-\frac{1}{\eta^2}
\left(\partial_t+\vau_0\cdot\na\right)\delta\hat S
\,.
\ea
Hence $\delta\hat\varrho$ is indeed the (negative) canonical conjugated 
momentum to $\delta\hat S$ -- provided that one inserts the constant factor 
$\eta^2$ correctly into the (effective) action -- and the commutation relations
are equivalent (within the used approximation) to those of a quantum 
field in a curved space-time.

\subsection{Non-Dispersive Dielectric Media}
\label{Non-dispersive Dielectric Media}

As another example we study non-dispersive and linear dielectric media,
see e.g.~\cite{DBHA}.
For a medium at rest the fundamental Lagrangian describing the 
electromagnetic field, the dynamics of the medium (${\cal L}[\f{P}]$),
as well as their mutual interaction ($\f{E}\cdot\f{P}$) is given by 
\bea
{\cal L}=\frac{1}{2}\left(\f{E}^2-\f{B}^2\right)+\f{E}\cdot\f{P}+
{\cal L}[\f{P}]
\,.
\ea
Accordingly, using the temporal gauge and introducing the vector potential
via $\f{E}=\partial_t{\f{A}}$ and $\f{B}=-\na\times\f{A}$, 
the canonical momentum is just the electric displacement
\bea
\f{\Pi}=\f{D}=\f{E}+\f{P}
\,.
\ea
Performing basically the same steps as in Sec.~\ref{Effective Action} we may 
integrate out the degrees of freedom associated to the medium $\f{P}$ and 
thereby arrive at an effective (low-energy) action for the (macroscopic)
electromagnetic field alone, cf.~\cite{DBHA}.
But, in contrast to the highly resonant behavior of $\f{P}$ 
in slow-light systems, non-dispersive media respond adiabatically
with a constant susceptibility $\chi=\varepsilon-1$,
i.e., $\f{P}=\chi\f{E}$ and thus $\f{\Pi}=\f{D}=\varepsilon\f{E}$,
to the external field (at sufficiently low frequencies), cf.~\cite{DBHA}.

If the (non-dispersive) medium is moving with the velocity $\f{\beta}$
the electric and magnetic fields get mixed and one obtains
\bea
\f{\Pi}=\f{D}=\varepsilon\f{E}+(\varepsilon-1)\f{B}\times\f{\beta}
+\ord(\f{\beta}^2)
\,.
\ea
Again, the commutation relations fit to an effective-metric description 
-- which is not completely surprising because the effective action has the 
same form as in curved space-times, cf.~\cite{DBHA}.

\section{Discussion}\label{Discussion}

Let us summarize:
The naive application of slow light (i.e., the most common set-up) in order to 
create a black hole analog goes along with several problems, 
cf.~Sec.~\ref{Problems of Slow Light}.
With the scenario proposed in this article, the problems associated to the
classical wave equation can be solved and it is -- at least in principle -- 
possible to create a (classical) black hole analog for the $\Phi$ field.
At low wave-number, the corresponding dispersion relation represents a 
quadratic relation between $\kappa$ and $\omega$, and can thus be written 
in terms of an effective metric. 
If the fluid is in motion, this low wave-number equation can be changed into 
a black hole type wave equation.

However, this classical black hole analog does {\em not} reproduce the 
expected quantum effects -- such as Hawking 
radiation\footnote{
This conclusion applies in the same way to the scenario proposed in 
Ref.~\cite{leo-nature}, where the Schwarzschild metric is simulated by a 
medium at rest with the horizon corresponding to a singularity in the 
effective refractive index.
Such static analogs of the Schwarzschild geometry (see also \cite{static})
go along with further problems \cite{on-static}.}.
In order to simulate the Hawking effect, it is not sufficient to 
design a system with an equivalent effective equation of motion -- 
the commutation relations have to match as well.
This is indeed the case for the sonic black hole analogs in Bose-Einstein 
condensates and non-dispersive dielectric black hole analogs -- 
but for sound waves in more complicated systems, for example, 
it is not immediately obvious.

Nevertheless, in the scenario described in this article, the field $\Phi$ 
governing the beat fluctuations of an electromagnetic background field obeys 
the same equation of motion as in the presence of a horizon and hence
can be used to model several classical effects associated to black holes -- 
for example the mode mixing at the horizon as described by the Bogoliubov 
coefficients, see Sec.~\ref{Bogoliubov-Dirac}.
One way of measuring the Bogoliubov coefficients could be to send in a 
``classical'' pulse above the background -- i.e., a particular coherent sate 
in terms of the fundamental electromagnetic field -- and compare it with the 
outcoming pulse.
As another (more fancy) possibility one might think of a multi-mode 
squeezed state -- which in some sense simulates the vacuum fluctuations 
which are transformed into quasi-particles by the mode mixing.

However, one should bear in mind that, as the wave-packets propagate away from
the horizon and get strongly blue-shifted, they eventually reach the regime
where the concept of the effective geometry breaks down and effects like 
dispersion, non-locality (in time) of the effective action, and, finally, 
absorption become relevant.
For a reasonably clean interpretation, therefore, one should investigate the
scattering of the wave-packets not too far away from the horizon.

\subsection{Miles Instability}\label{Miles instability}

Another interesting classical effect is related to the negative parts of the
energy in Eq.~(\ref{negative}).
Since a conserved positive definite energy functional of the linearized 
perturbations would demonstrate linear stability, the negative contribution 
in Eq.~(\ref{negative}) can be interpreted as an indicator for a potential 
instability (e.g., super-radiance) -- provided a suitable coupling between 
positive and negative (effective) energy modes.

As an example, let us assume that the ``superluminally'' flowing 
$v>1/\sqrt{1+2\aleph}$ slow-light medium interacts with the environment 
in the laboratory frame via a friction term such as $\Gamma\partial_t\Phi$ 
(with possible spatial derivatives).
For small $\omega$ and $\kappa$ the resulting dissipation alters the
dispersion relation via
\bea
\left(\omega+v\kappa\right)^2=\frac{\kappa^2}{1+2\aleph}-i\omega\Gamma(\kappa)
\,,
\ea
with the potentially $\kappa$-dependent (additional spatial derivatives) 
friction term $\Gamma(\kappa)$ describing the interaction of the $\Phi$-field 
with the environment at rest.
 
For small $\Gamma$ the imaginary part of the solutions for the frequency 
$\omega$ (assuming a real wave-number $\kappa\in{\mathbb R}$) reads
\bea
\Im(\omega)=-\frac{\Gamma(\kappa)}{2}
\left(1\pm v\sqrt{1+2\aleph}\right)
\,.
\ea
Consequently, beyond the horizon $v>1/\sqrt{1+2\aleph}$ one of the allowed
frequency solutions acquires a positive imaginary part and thus the
dissipation (interaction with the environment) generates an instability.
Note that the relative velocity $v>1/\sqrt{1+2\aleph}$ between the slow-light 
medium and the environment (at rest) is crucial since a friction term like
$\Gamma(\partial_t+v\partial_x)\Phi \to i(\omega+v\kappa)\Gamma$ 
would of course not lead to any instability.

This instability is somewhat analogous to the Miles instability \cite{miles} 
generating surface waves in water by wind blowing over it.
In Ref.~\cite{volovik}, this phenomenon is called thermodynamic instability
since it occurs when the free energy of the medium acquires negative parts
in the frame of the environment.

\subsection{Outlook}\label{Outlook}

Apart from the aforementioned experiments there are many more conceivable 
tests one could perform with the proposed classical black hole analog based 
on slow light.
A more drastic way of investigating the interior structure of the sample
(than the mere comparison of the in- and out-states) could be to 
freeze the dark state by completely switching off the background field
and take a ``snap-shot'' of the state of the atoms by illuminating them 
with strong laser beams with frequencies corresponding to certain atomic 
transitions and measuring the absorption.

Furthermore it would be interesting to investigate the influence of the 
anomalous frequency solutions of the dispersion relation generated by the 
non-local temporal dependence (cf.~Sec.~\ref{Dispersion relation}),
for example, on additional mode-mixing. 
This question is relevant for more general (non-dispersive) dielectric 
black hole analogs and might also lead to some insight into the 
trans-Planckian problem.

\section*{Acknowledgements} 

W.~G.~U.~would like to thank the Canadian Institute for Advanced Research. 
R.~S.~would like to thank the Alexander von Humbolt foundation for support, 
and both also thank the Natural Science and Engineering Research Council of 
Canada for support.


\addcontentsline{toc}{section}{References}

\end{document}